\documentclass[twocolumn]{aastex631}

\usepackage{color}
\usepackage{url}
\shorttitle{X-ray polarization in PG~1553+113}
\shortauthors{Middei et al.}

\newcommand\xmm{\rm{XMM-Newton~}}
\newcommand\swift{\rm{Swift}}
\newcommand\ixpe{\rm{IXPE}}

\begin{document}

\title{X-ray polarimetry of PG~1553+113}
\title{IXPE and multi-wavelength observations of blazar PG~1553+113 reveal an orphan optical polarization swing}

%\correspondingauthor{August Muench}
%\email{greg.schwarz@aas.org, gus.muench@aas.org}

%%%%%%%%%%%%%%%%%%%%%%%

%
%
%
%TIER 1
%
%
%

%--------------------------------main contributors

\author[0000-0001-9815-9092]{Riccardo Middei}
\correspondingauthor{Riccardo Middei}
\email{riccardo.middei@ssdc.asi.it}
\affiliation{Space Science Data Center, Agenzia Spaziale Italiana, Via del Politecnico snc, 00133 Roma, Italy}
\affiliation{INAF Osservatorio Astronomico di Roma, Via Frascati 33, 00078 Monte Porzio Catone (RM), Italy}

\author[0000-0000-0000-0000]{Matteo Perri}
\affiliation{Space Science Data Center, Agenzia Spaziale Italiana, Via del Politecnico snc, 00133 Roma, Italy}
\affiliation{INAF Osservatorio Astronomico di Roma, Via Frascati 33, 00078 Monte Porzio Catone (RM), Italy}

\author[0000-0000-0000-0000]{Simonetta Puccetti}
\affiliation{Space Science Data Center, Agenzia Spaziale Italiana, Via del Politecnico snc, 00133 Roma, Italy}

\author[0000-0001-9200-4006]{Ioannis Liodakis}
\affiliation{Finnish Centre for Astronomy with ESO, FI-20014 University of Turku, Finland}

\author[0000-0000-0000-0000]{Laura Di Gesu}
\affiliation{ASI - Agenzia Spaziale Italiana, Via del Politecnico snc, 00133 Roma, Italy}

\author[0000-0001-7396-3332]{Alan P. Marscher}
\affiliation{Institute for Astrophysical Research, Boston University, 725 Commonwealth Avenue, Boston, MA 02215, USA}

\author[0000-0001-7396-3332]{Nicole Rodriguez Cavero}
\affiliation{Physics Department, McDonnell Center for the Space Sciences, and Center for Quantum Leaps, Washington University in St. Louis, St. Louis, MO 63130, USA}

\author[0000-0003-0256-0995]{Fabrizio Tavecchio}
\affiliation{INAF Osservatorio Astronomico di Brera, Via E. Bianchi 46, 23807 Merate (LC), Italy}

\author[0000-0002-4700-4549]{Immacolata Donnarumma}
\affiliation{ASI - Agenzia Spaziale Italiana, Via del Politecnico snc, 00133 Roma, Italy}

\author[0000-0002-4700-4549]{Marco Laurenti}
\affiliation{INFN - Sezione di Roma “Tor Vergata”, Via della Ricerca Scientifica 1, 00133 Roma, Italy}
\affiliation{Space Science Data Center, SSDC, ASI, Via del Politecnico snc, 00133 Roma, Italy
}

%% Multiwavelength contributors + TWG----------------
% ...

\begin{abstract}

The lower energy peak of the spectral energy distribution of blazars has commonly been ascribed to synchrotron radiation from relativistic particles in the jets. Despite the consensus regarding jet emission processes, the particle acceleration mechanism is still debated. Here, we present the first X-ray polarization observations of PG~1553+113, a high-synchrotron-peak blazar observed by the Imaging X-ray Polarimetry Explorer (\ixpe). We detect an X-ray polarization degree of $(10\pm2)\%$ along an electric-vector position angle of $\psi_X=86^{\circ}\pm8^{\circ}$. At the same time, the radio and optical polarization degrees are  lower by a factor of $\sim$3. During our \ixpe\ pointing, we observed the first orphan optical polarization swing of the \ixpe~era, as the optical angle of PG 1553+113 underwent a smooth monotonic rotation by about 125$^\circ$, with a rate of $\sim$17 degree per day. We do not find evidence of a similar rotation in either radio or X-rays, which suggests that the X-ray and optically emitting regions are separate or, at most, partially co-spatial. Our spectro-polarimetric results provide further evidence that the steady-state X-ray emission in blazars originates in a shock-accelerated and energy-stratified electron population.
\end{abstract}


\section{Introduction} \label{sec:intro}

Blazars, a rare class of active galactic nuclei \citep[AGN;][]{Antonucci1993,Urry1995,Padovani2017}, display highly time-variable radiation  across the entire electromagnetic spectrum, from radio to very high-energy (TeV) $\gamma$-rays \citep[e.g.,][]{Angel1980,Padovani1995,Ghisellini1998,Ghisellini1998b,Sreekumar1998,Fossati1998,Abdo2010,Maselli2010,Acero2015,Agudo2011,Agudo2011bis,Liodakis2018,Liodakis2019-II,Hovatta2019}.
It is widely accepted that the nonthermal radiation originates in a highly collimated jet of relativistic particles extended along the polar axis of an accreting supermassive black hole (SMBH) and pointing toward the Earth \citep[see e.g.][]{Blandford2019}. Despite accretion of matter onto the SMBH powering blazars, their orientation is such that highly relativistically boosted radiation from the jet dominates the emission spectrum \citep[e.g.,][]{Laahteenmaki1999,Hovatta2009,Liodakis2018-II}, with a double-peaked spectral energy distribution (SED). The first hump peaks between infrared (IR) and X-ray frequencies, and is ascribed to synchrotron radiation from energetic electrons, perhaps, as suggested by previous \ixpe~pointings, accelerated in shocks \citep[e.g.,][]{Liodakis2022-Mrk501,DiGesu2022-Mrk421}. The second hump is located in the GeV -- TeV range, with the mechanism behind the emission still debated. Both hadronic and leptonic frameworks have been proposed \citep[e.g.][]{Maraschi1992,Boettcher2012,Cerruti2015}, although recent X-ray spectro-polarimetric observations seem to favor a leptonic scenario in which synchrotron self-Compton emission plays a key role in shaping the SED of some blazars \citep[see e.g.][]{Middei2023,Peirson2023}. Depending on the frequency at which the synchrotron component peaks ($\nu_{peak}$), blazars are classified as high-synchrotron-peak (HSP), intermediate-synchrotron-peak (ISP) or low-synchrotron-peak (LSP,) with $\nu_{\rm peak}>$10$^{15}$ Hz, $10^{14}<\nu_{\rm peak}<10^{15}$ Hz, and $\nu_{\rm peak}<10^{14}$ Hz, respectively \citep[e.g.,][]{Ajello2020}.\\

 PG~1553+113 (RA=15h 55m 43.0440s, Dec$=+11^{\circ}$ $11'$ $24.365''$, J2000) is an HSP source ($\nu_{\rm peak}\approx3.9\times10^{15}$ Hz, \citealp{Ajello2020}), therefore X-rays sample the falling part of the low-energy hump of its SED. Although several redshift values ($z<$0.5) have been reported for this BL Lacertae object, the most commonly adopted value is $z$=0.433 \citep[see, e.g.,][]{Nicastro2018,Johnson2019,Dorigo2022}.  It is a TeV $\gamma$-ray emitter \citep{Aleksic2012}, and has exhibited one of the most compelling cases of quasi-periodic oscillations from a blazar at $\gamma$-ray energies as a period p=2.18$\pm$0.08 (years) was found in the $\gamma$-rays \citep{Ackermann2015,Tavani2018}.

Here we present the first X-ray polarization observation of PG 1553+113, an HSP blazar observed by the Imaging X-ray Polarimetry Explorer \cite[\ixpe;][]{Weisskopf2022} in 2023 February.
Owing to the three gas pixel detectors \cite[GPDs;][]{Costa2001,Bellazzini2003,Baldini2021} in its focal plane, \ixpe\ provides an unprecedented opportunity to measure the X-ray polarization of cosmic sources. Since its launch on 2021 December 9, \ixpe\ has observed different types of AGNs \citep[e.g.,][]{Marinucci2022,Ursini2023,Gianolli2023,Ingram2023arXiv,Tagliacozzo2023arXiv,Ehlert2022}, including blazars \citep[e.g.][]{Liodakis2022-Mrk501,DiGesu2022-Mrk421,Middei2023}.
PG~1553+113 is one of several HSP blazars observed by \ixpe.
%, and, among these, it has the second lowest synchrotron peak.
In \S\ref{sec:xray+pol} we report on the data reduction, while \S\ref{sec:xray_analysis} discusses the X-ray emission and polarization properties of PG 1553+113. In \S\ref{sec:Multi_data} we present the contemporaneous radio and optical polarization observations, and in \S\ref{sec:disc_conc} we summarize and discuss our results and draw conclusions.

\section{X-ray data reduction}\label{sec:xray+pol}

		\begin{table}
		\centering
		\caption{\small{Log of the \ixpe~ and \xmm observations of PG 1553+113.}}\label{log}
		\begin{tabular}{c c c c}
			\hline
			Observatory & Obs. ID & Obs. date & Net exp. \\
			& & yyyy-mm-dd/dd & ks \\
			\hline
			\ixpe (a) & 02004999  & 2023-02-01/03 &$\sim$ 98\\
   			\ixpe (b) & 02004999  & 2023-02-08/09 &$\sim$ 28\\
         	\xmm & 0902112101& 2023-01-31&$\sim$14\\
            \hline
		\end{tabular}
	\end{table}

PG 1553+113 was observed with \ixpe's three Detector Units (DUs)  in 2023 February for a total exposure time of $\sim$130 ks. The observing time was split into two intervals (hereafter, intervals a and b). Interval a was taken in 2023 February 1-2, for a total net exposure time of $\sim$100 ks. Interval b, performed one week later in February 8-9, had a net exposure time of $\sim$30 ks.

We coordinated the \ixpe\ observations with {\rm XMM-Newton} \citep[][]{Jansen2001} and the Neil Gehrels Swift observatory \citep[Swift;][]{Gehrels2004}, which observed the target before and after the \ixpe\  observation. Table~\ref{log} reports the log for the \ixpe\ and \xmm\ pointings.

We extracted the $I$, $Q$ and $U$ spectra for each of the three DUs of \ixpe\ using the software \textsc{ixpeobssim} \citep[v. 30.0.0;][]{pesce2019,baldini2022} and following the background rejection prescriptions presented in \citet{DiMarco2023}. Moreover, spectra were computed in order to use the so-called weighted analysis method presented in \citet[][parameter STOKES = NEFF in XSELECT]{dimarco2022}. The spectra for the Stokes parameters were computed from a circular region with radius=0.95$'$ centered on the source, while an annulus with $r_{\rm in(out)}$=1.2$'$(3.5$'$) was adopted to extract the background. This approach has been shown to enhance the sensitivity to polarization \citep{DiMarco2023}. The resulting $I$ Stokes spectra were grouped by requiring each energy bin to have a signal-to-noise ratio greater than 7. We adopted uniform binning of 280 eV for the $Q$ and $U$ Stokes spectra.

\xmm\ performed a snapshot ($\sim$14 ks long) of PG 1553+113, and we analyzed the data taken with the EPIC-pn camera \citep[][]{Struder2001}. The source was observed in Small Window mode, with the medium filter applied. We processed the data via the standard {\rm XMM-Newton} Science Analysis System
(SAS v21, \citealp{Gabriel2004}). Source extraction radii and screening for high-background intervals were determined through an iterative process \citep{Piconcelli2004} that maximizes the signal-to-noise ratio. The background was extracted from circular regions with a radius of 50$''$, and the same shape centered on the source PG~1553+113 was adopted for science products. The resulting third-level products were grouped by requiring each bin to contain at least 30 counts, and not to over-sample the spectral resolution by a factor larger than 3. The net count rate was less than the maximum allowed limit of 50 cts s$^{-1}$ to avoid deteriorated response due to photon pile-up for EPIC-pn observations in Small Window mode \citep[e.g.][]{Jethwa2015}. We further assessed the potential impact of pile-up in the \xmm observation by means of the \emph{epatplot} task, a standard SAS command devoted to checking for any pile-up affecting the data, and we found it to be negligible.

\swift\ pointed at PG~1553+113 before and after the \ixpe\ pointing. Sixteen exposures, each 1 ks long, were taken in photon counting mode and calibrated and reduced using the {\it XRTDAS} software package\footnote{\url{https://sda2006.ts.infn.it/presentazioni/capalbi.pdf}}. To extract the source spectra, we adopted a circular region of radius $47''$, while the background was derived with a concentric annulus with inner(outer) radii of $120''$($150''$). Spectra were then binned, with at least 25 counts in each bin, in order to meaningfully use the $\chi^2$ statistic in our spectral analysis.

\section{X-ray analysis}\label{sec:xray_analysis}
The X-ray spectro-polarimetric analysis was performed using \textsc{XSPEC} \citep[][]{Arnaud1996} and following the prescriptions of the so-called weighted analysis method \citep[][]{dimarco2022}. We accounted for Galactic absorption along the line of sight using the {\tt tbabs} model. In all the fits, we set the Galactic column density to n$_{\rm H}$=3.62$\times10^{20}$ cm$^{-2}$ \citep{HI4PI2016}. For calculations involved the distance to PG~1553+113, we adopt the standard cosmological framework with $H_0$ = 70 km s$^{-1}$ Mpc$^{-1}$
%, $q_0$ = 0.0,
and $\Lambda_0$ = 0.73.

\subsection{\ixpe\ and \xmm\ spectro-polarimetric analysis}
To determine the spectral properties of PG~1553+113, we modeled the 0.3--10 keV \xmm\  spectrum (taken contemporaneously with the \ixpe\ observation; see Table~\ref{log}). We tested two models: an absorbed power law and an absorbed logarithmic parabola. The log-parabola \citep{Massaro2004} is defined by
$N(E)=(E/E_1)^{-\alpha-\beta\log(E/E_1)}$,
where $\alpha$ accounts for the source photon index at energy E$_1$, $\beta$ represents the curvature of the parabola, and E$_1$ is a reference (``pivot'') energy.
In the fits, we determined the photon index and normalization for the power law. For the log-parabola, we fixed E$_1$ to 3 keV and left $\alpha$, $\beta$, and the normalization free to vary. The power law fails to reproduce the \xmm spectrum satisfactorily ($\chi^2$/d.o.f.=530/159), while the logarithmic parabola returns a better statistic, with $\chi^2$=219 for 159 degrees of freedom. We obtained $\alpha=2.50\pm0.01$ and $\beta=0.13\pm 0.01$, which is compatible with a soft and curved spectrum, as expected for a blazar X-ray spectrum dominated by synchrotron radiation. We note that this fit can be further improved ($\Delta\chi^2$=--41 for two additional degrees of freedom) by adding a narrow Gaussian component to account for an additional soft X-ray feature ($E=0.68$ keV and Norm=2.58 $\times$10$^{-4}$ photons cm$^{-2}$ s$^{-1}$).
Finally, we notice that \ixpe\ caught PG 1553+113 in a fairly high flux state compared with the average flux of this source ($F_{\rm mean}\sim$2.1 $\times$ 10$^{-11}$ erg cm$^{-2}$ s$^{-1}$ in the 2--10 keV energy range) obtained by \citet{Giommi2021} who analysed 802 {\rm Swift} archival snapshots of the blazar.

\begin{figure*}
    \centering
    \includegraphics[width=\textwidth]{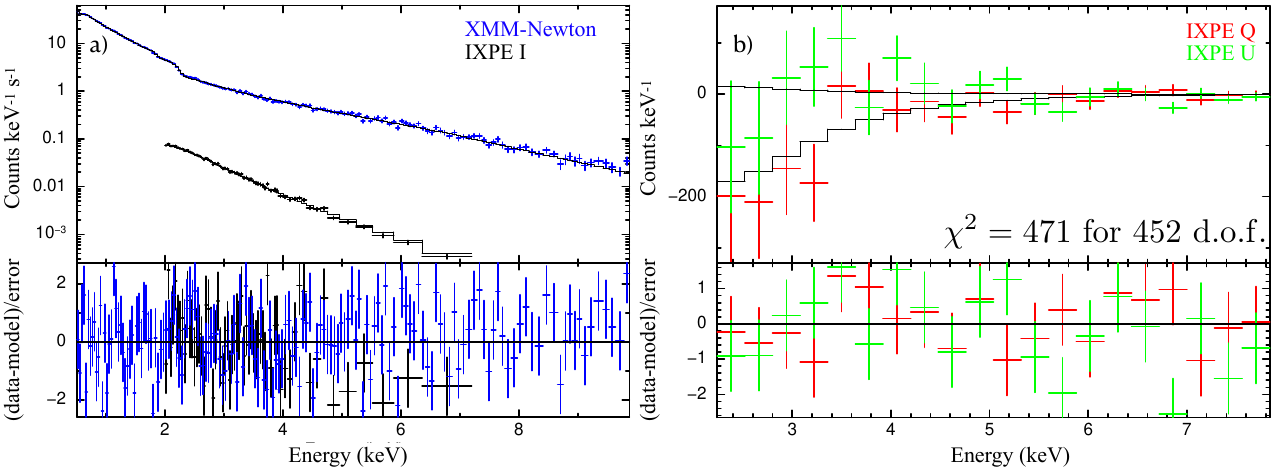}
    \includegraphics[width=0.99\textwidth]{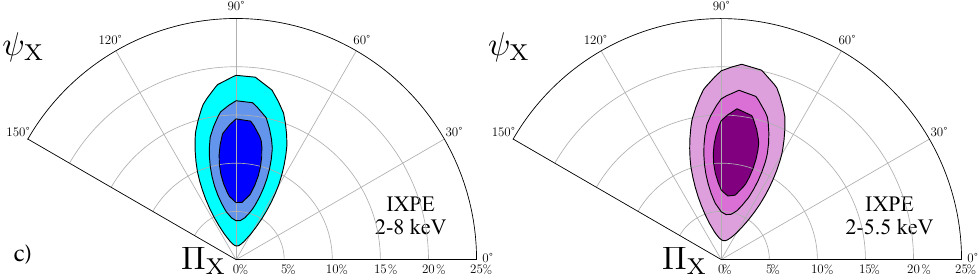}
    \caption{Top Panels: Best fit to the \ixpe\ and \xmm\ data and corresponding residuals. On the left, we show the model \texttt{const$\times$polconst$\times$logpar} fitting the $I$ Stokes spectra, while on the right the best-fit $Q$ and $U$ Stokes spectra are shown. Bottom panels:  regions accounting for 68\%, 90\%, and 99\% confidence levels computed using the full \ixpe\ 2-8 keV data (bottom left) or taking into account the \ixpe\ $I$, $Q$ and $U$ Stokes spectra only in the 2-5.5 keV range.}
    \label{xspecpol}
\end{figure*}

Next, we included in the analysis the \ixpe\ spectra extracted from both the a and b intervals. We assigned each spectrum to a ``datagroup'' in \textsc{XSPEC}, which allowed us to couple/decouple the parameters of the model between the different instruments when needed. As a first test, we assumed the polarization properties of PG 1553+113 to be the same in the two observing intervals. Thus, we fit the corresponding $I$, $Q$, and $U$ Stokes spectra while constraining the relevant parameters. To account for polarization, we updated our \textsc{XSPEC} model to {\tt tbabs$\times$const$\times$polconst$\times$logpar}.
The constant model (\texttt{const}) allows for cross-calibration among the different \ixpe\ DUs and the EPIC-pn camera. The polarization model {\tt polconst}, which assumes constant polarization parameters within the operating energy range, has two free parameters: the polarization degree ($\Pi_{\rm X}$) and the electric-vector position angle ($\Psi_{\rm X}$, measured from north through east). Finally, for the \texttt{logpar} model, we coupled the $\alpha$ and $\beta$ parameters between the \ixpe\ and \xmm\ spectra. This model provided a statistically acceptable ($\rm \chi^2$ / d.o.f.=471/452) fit to the \ixpe\ and \xmm\ data, which is shown in Fig.~\ref{xspecpol} (panels a and b for the $I$ and $Q$ and $U$ Stokes parameters, respectively). The corresponding best-fit parameters are listed in Table~\ref{best-fitTable}.
Our fit result suggests that the spectrum of PG~1553+113 had a soft and curved shape ($\alpha=2.49\pm$0.01 and $\beta=0.11\pm$0.01). These values are well within the range of what is commonly observed in HSP blazars \citep[e.g.][]{Middei2022blaz}. The polarization properties were determined to be $\Pi_X$=(10.1$\pm$2.3)\% and $\psi_X$ =86$^{\circ}\pm8^{\circ}$. In Fig.~\ref{xspecpol}c we show the contours corresponding to 68\%, 90\%, and 99\% confidence levels. We find the polarized signal to be maximum in terms of significance in the 2--5.5 keV energy range. The soft and curved shape of the spectrum, and the maximum polarization signal measured below 5.5 keV, are in agreement with the PG~1553+113 X-ray spectrum being dominated by synchrotron emission.

\indent We further searched for a possible change in the polarization properties of PG~1553+113 between observation intervals a and b. Using the model {\tt tbabs$\times$const$\times$polconst$\times$logpar}, we computed $\alpha$, the normalization, $\Pi_X$, and $\psi_X$ from the Stokes parameters derived for intervals a and b separately. In this test the curvature parameter $\beta$ was kept frozen to its best-fit value indicated in Table~\ref{best-fitTable}\footnote{For this test we did not considered the \xmm data, as they were taken just before the \ixpe\ interval a exposure}. The adoption of the {\tt tbabs$\times$const$\times$polconst$\times$logpar} model led us to a very good representation of the two data sets, with $\chi^2$/d.o.f.=304/276 and $\chi^2$/d.o.f.=285/276 for intervals a and b, respectively. The inferred best fit parameters are listed in Table \ref{best-fitTable}. Despite the compatible 2-8 keV flux observed in the two intervals, $\Pi_{\rm X}$ and $\psi_{\rm X}$ are only constrained within interval a, while an upper limit $\Pi_{\rm X}^{99\%}<$26\% is obtained when fitting the Stokes $I$, $Q$ and $U$ spectra from interval b. This upper limit is likely the result of the relatively short duration of the b interval exposure.
As a final test, we studied the polarization properties of PG 1553+113 over time spans shorter than the duration of intervals a and b of the \ixpe~observation. However, for all the tested temporal intervals no hints of variability were found.

\begin{table}
\setlength{\tabcolsep}{.9pt}
\centering
	\caption{Best-fit parameters from the spectro-polarimetric analysis of PG 1553+113. Time-average results refer to the \ixpe~and \xmm joint fit, while values quoted for intervals a and b were obtained only with the \ixpe~data. Fluxes are in units of $10^{-11}$ erg cm$^{-2}$ s$^{-1}$ while the logarithmic parabola has a normalization of $10^{-3}$. All errors are given at 68\% confidence interval for 1 parameter of interest (i.e. $\Delta \chi^2$=1), while the upper limit to the polarization degree corresponds to 99\% uncertainty. The symbol $\dagger$ indicates those parameters that were kept frozen during the model fits. \label{best-fitTable}}
	\begin{tabular}{c c c c c}
	\hline
	Model & Parameter & time average & Interval a & Interval b \\
	\hline
    polconst& $\rm \Pi_X$ &(10.1$\pm$2.3)\% &(11.6$\pm$3.4)\%&$<26\%$ \\
     &  $\psi_X$ & 86$^{\circ}\pm8^{\circ}$& 85$^{\circ}\pm8^{\circ}$&-\\
    tbabs& n$_{\rm H}$ &3.62$\dagger$ &3.62$\dagger$&3.62$\dagger$\\

    log-par& $\alpha$ & 2.491$\pm0.007$& 2.59$\pm$0.03 &2.56$\pm$0.15\\
    & $\beta$ & 0.11$\pm0.01$&0.11$\dagger$&0.11$\dagger$ \\
    & Norm &$1.49\pm0.02$ &1.41$\pm$0.03&1.43$\pm$0.02 \\
    const & K$_{\rm Du2}$ & 0.91$\pm$0.01&0.95$\pm$0.01&0.96$\pm$0.07\\
    const & K$_{\rm Du3}$ &0.95$\pm$0.01 &0.90$\pm$0.01&0.97$\pm$0.07\\
    const & K$_{\rm Epic-pn}$ & 1.04$\pm$0.01&-&-\\
    \hline
    $F_{\rm 2-8~keV}$&&2.55$\pm$0.03&2.47$\pm$0.03&2.44$\pm$0.05 \\
    \hline
	\end{tabular}
	\end{table}

\subsection{{\rm Swift} monitoring campaign}

\begin{figure*}
    \centering
    \includegraphics[width=0.99\textwidth]{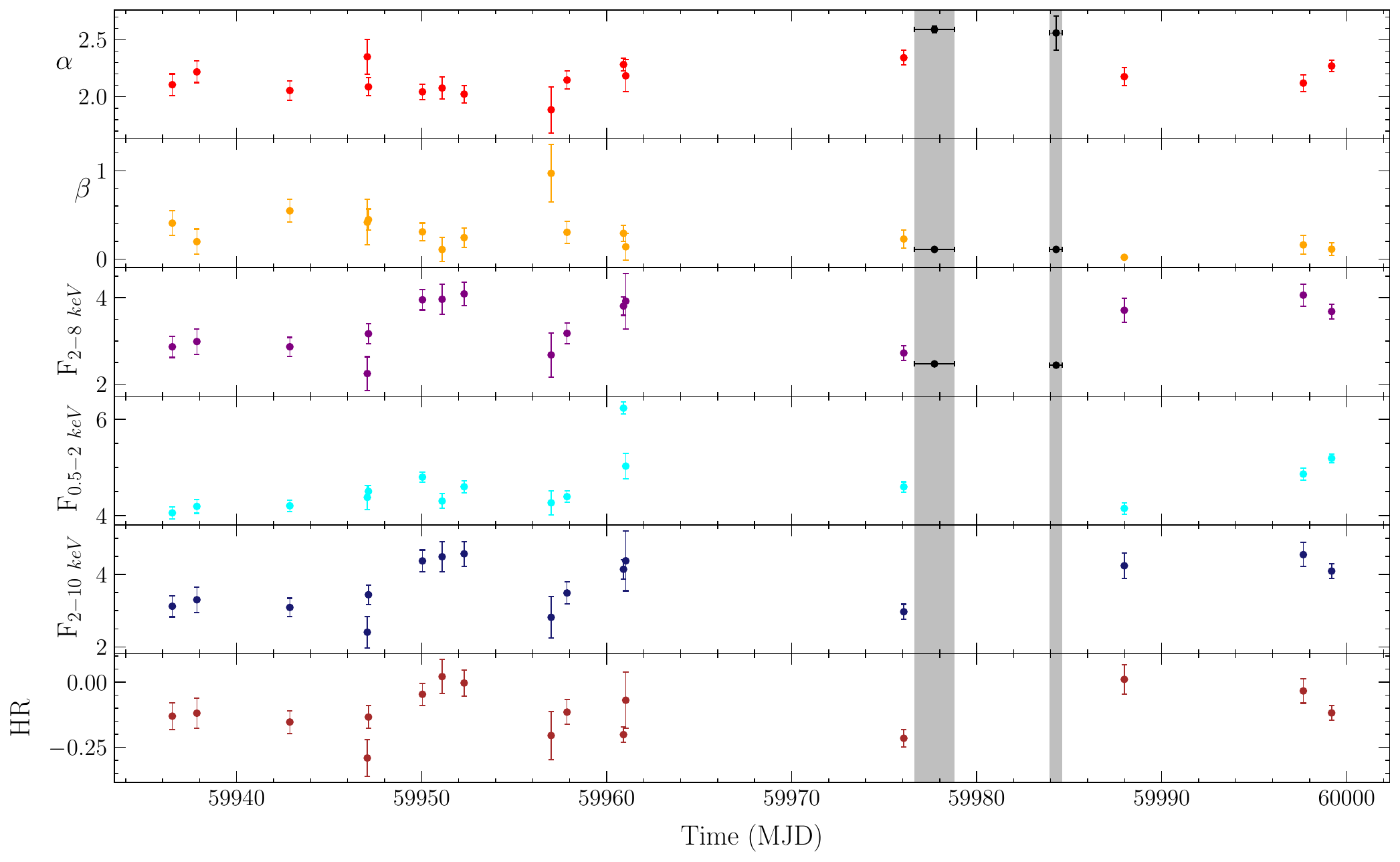}
    \caption{Results of Swift-XRT monitoring campaign of PG~1553+113. Fluxes F$_{\rm IXPE}$, F$_{\rm Soft}$, and F$_{\rm Hard}$ are computed in the 2-8, 0.5-2, and 2-10 keV bands, respectively, and are displayed in units of $10^{-11}$ erg cm$^{-2}$ s$^{-1}$. ``HR'' is the hardness ratio, defined as (F$_{\rm Hard}$-F$_{\rm Soft}$)/(F$_{\rm Hard}$+F$_{\rm Soft}$).Gray shaded areas mark the duration of the two \ixpe\ observations, intervals a (left) and b (right). Black data points account for the different quantities  measured by \ixpe~and quoted in Table~\ref{best-fitTable}.}
    \label{swiftPG1553}
\end{figure*}

A {\rm Swift} campaign was organized to monitor the flux and spectral variability of PG~1553+113. The {\it XRT} observed the blazar from 2023 early January until late February. We fit the resulting 16 spectra (exposures of $\leq$ 1 ks each) with a log-parabola model, with Galactic absorption (see above). The adopted model provides a statistically acceptable representation of the data. We show the spectral parameters and flux estimates inferred from the fits in Fig.~\ref{swiftPG1553}. The
{\rm Swift} data are compatible with nearly constant spectral shape and variability of flux by less than a factor of 2. The average values for the quantities displayed in Fig.~\ref{swiftPG1553} are: $\alpha_{\rm avg}$=2.15$\pm$0.12 and $\beta_{\rm avg}$=0.3 $\pm$0.2, with fluxes F$^{\rm avg}_{\rm IXPE}=3.4\pm$0.6 (2-8 keV), F$^{\rm avg}_{\rm Soft}=4.6\pm0.5$ (0.5-2 keV), and F$^{\rm avg}_{\rm Hard}=3.7\pm0.6$ (2-10 keV) in units of $10^{-11}$ erg cm$^{-2}$ s$^{-1}$.

\section{Multi-wavelength observations} \label{sec:Multi_data}

\begin{figure*}
    \centering
     \includegraphics[width=\textwidth]{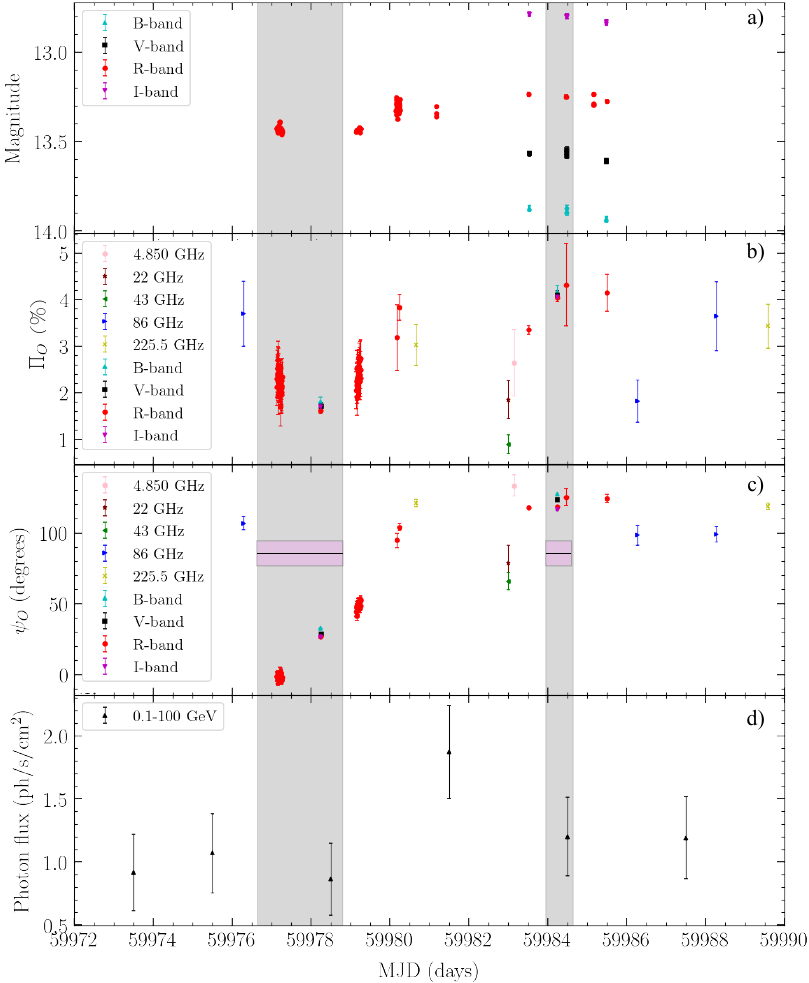}
    \caption{Multi-wavelength radio and optical observations of PG~1553+113. Panel a shows the brightness in magnitudes, panel b the polarization degree, panel c the polarization angle, and panel d the $\gamma$-ray light curve in the 0.1-100~GeV range. The gray shaded areas mark the two intervals of \ixpe\ observations. The error bars represent 1$\sigma$ uncertainties. Pink boxes identify the $\psi_X$ best-fit value and its 1-$\sigma$ uncertainty as derived using \ixpe. Panel d displays the photon flux observed by the \rm{FERMI~} Large Area Telescope during the \ixpe\ monitoring of the source. The photon flux is in units of $10^{-7}$ photons s$^{-1}$ cm$^{-2}$ \citep[see][for details of the {\rm Fermi} light curve]{Abdollahi2023}.}
    \label{plt:multi_obs}
\end{figure*}

PG~1553+113 was also observed at radio and optical wavelengths by the Effelsberg 100-m telescope, the Korean VLBI Network (KVN), Institut de Radioastronomie Millim\'etrique (IRAM 30-m telescope), Submillimeter Array (SMA), Boston University Perkins Telescope, Calar Alto Observatory, Nordic Optical Telescope and Sierra Nevada Observatory. The observational and data analysis procedures are described in detail in \citet{Middei2023}. All the multiwavelength observations are available upon request to the individual observatories. We supplement our campaign with 3-day-binned publicly available {\rm Fermi} Large Area Telescope data from the light curve repository\footnote{\url{https://fermi.gsfc.nasa.gov/ssc/data/access/lat/LightCurveRepository/}} \citep{Abdollahi2023}.

At radio wavelengths, the Effelsberg observation on 2023 February 8 (MJD~59983.14) was performed at 4.85~GHz as part of the QUIVER (Monitoring the Stokes $Q$, $U$, $I$ and V Emission of AGN jets in Radio) program \citep{Krauss2003,Myserlis2018}. The IRAM-30m observations were obtained and analyzed in the framework of the POLAMI (Polarimetric Monitoring of AGN at Millimeter Wavelengths) project\footnote{\url{http://polami.iaa.es/}} \citep{ Agudo2018-II,Agudo2018, Thum2018}.  The Korean VLBI Network (KVN) observation was conducted on 2023 February 8 (UT 20) with two 21-m antennas (KVN Yonsei and Tamna) combined in single-dish mode. The polarization observations were conducted in position-switching mode \citep{Kang2015} and calibrated using an unpolarized (Jupiter) calibrator and a polarized (3C286, \citealp{Agudo2012}) calibrator for the polarization degree, and the Crab nebula for the polarization angle (152$^\circ$; \citealp{Aumont2010}). The SMA observations were performed within the framework of the SMA Monitoring of AGNs with Polarization (SMAPOL) program \citep{Ho2004,Marrone2008,Primiani2016}. The radio observations covered a frequency range from 4.85 to 225.5~GHz. At the lowest radio frequency (4.85~GHz), the polarization degree was $\Pi_\mathrm{4.85GHz}=2.6\pm0.7\%$ along a polarization angle $\psi_\mathrm{4.85GHz}=133\pm7^\circ$. At intermediate frequencies (22~GHz, 43~GHz), the polarization degree was $\Pi_\mathrm{22-43GHz}=1-2\%$, with $\psi_\mathrm{22-43GHz}=65-80^\circ$. At the higher frequencies (86~GHz, 225~GHz), we measured $\Pi_\mathrm{86-225GHz}\sim3\%$ along $\psi_\mathrm{86-225GHz}=100-120^\circ$. We did not observe strong variability in either the degree or angle of polarization at any of the radio frequencies before, during, or after the \ixpe\ observations (Fig. \ref{plt:multi_obs}).

Optical observations were obtained at the Calar Alto Observatory, Nordic Optical Telescope, Perkins Telescope, and Sierra Nevada Observatory. At Calar Alto Observatory, we used the 2.2 m telescope and CAFOS (Calar Alto Faint Object Spectrograph). Observations at Sierra Nevada Observatory used the T90 telescope. The data for both observatories were taken in $R$-band and analyzed following standard polarimetric procedures. $B, V, R$, and $I$ band polarimetric observations were obtained with the Alhambra Faint Object Spectrograph and Camera (ALFOSC) at the Nordic Optical Telescope, and analyzed using Tuorla Observatory's semi-automatic data reduction pipeline \citep{Hovatta2016,Nilsson2018}. Additional photometric ($B,~V,~R, ~I$) and polarimetric ($R$-band) data were obtained at the Perkins Telescope (1.8 m) using the PRISM camera. During \ixpe\ interval a, we found $\Pi_O=2.2\pm0.4\%$. During interval b, we observed an increase of the polarization degree to $\Pi_O=4.2\pm0.5\%$. At the same time, we observed a rotation of the polarization angle of about $\Delta\psi_R\approx125^\circ$ at a rate of about 17$^\circ$ day$^{-1}$ (see Fig. \ref{plt:multi_obs}, panel c). The duration, amplitude and rotation rate are within typical rotation parameters of other blazars \citep[e.g.,][]{Blinov2018}. Compared to the two previous rotations of the optical polarization vector of PG~1553+113 detected by the RoboPol program \citep{Blinov2015,Blinov2016,Blinov2018}, the event reported here has a similar amplitude, but a factor of 2-3 higher rotation rate. However, we note that, while the end of the rotation is apparent in our data, the rotation may have started before the beginning of our observations. It is therefore possible that we are underestimating the total amplitude and duration of the event.

\section{Discussion and Conclusions}\label{sec:disc_conc}

We have reported the first \ixpe\ observations of the HSP blazar PG~1553+113, finding an X-ray polarization degree $\Pi_{\rm X}$=(10.1$\pm$2.3)\% along an electric-vector position angle of $\psi_X$=86$^\circ\pm8^\circ$. These values were obtained by performing a spectro-polarimetric analysis, using \textsc{XSPEC}, of the $I$, $Q$, and $U$ Stokes spectra that were fit  with a quasi-simultaneous \xmm\ spectrum. The source spectrum can be fit with a log-parabola, which models the curved X-ray spectrum typical of synchrotron-dominated HSP blazars \citep[e.g.][]{Massaro2004,Giommi2021,Middei2022blaz}.

Our contemporaneous radio and optical polarization observations find a 2.5--5 times lower polarization degree, with a polarization angle roughly aligned with the X-ray value. Previous \ixpe\ results for the HSP blazars Mrk~421 and Mrk~501 \citep{Liodakis2022-Mrk501,DiGesu2022-Mrk421} have found similar chromatic behavior of $\Pi$, usually with similar $\psi$ across different frequencies. The ratio $\Pi_{\rm X}/\Pi_{\rm O}$ is also similar to that found in Mrk~421 and Mrk~501. These results have been interpreted as evidence of an electron population that is accelerated in a shock and becomes energy stratified as it cools while propagating away from the shock front \citep[see details in, e.g.,][]{Marscher1985,Perlman1999,Perlman2005,Angelakis2016,Tavecchio2018}. Given the different synchrotron cooling lengths, the optical and radio emitting particles occupy an increasing volume of the jet with decreasing frequency. In this case, the radio, optical, and X-ray emission regions are expected to be at most partially co-spatial, which leads to different polarization properties in different bands.

Rotations with time of the optical polarization angle have now been observed in a large number of blazars \citep[e.g.,][]{Marscher2008,Marscher2010,Blinov2016-II,Blinov2018,Liodakis2020}. Recently, rotation of the X-ray polarization vector was observed for the first time, in Mrk~421 \citep[][]{DiGesu2023}. During this event, Mrk~421 was observed to have a variable X-ray spectral shape and flux, but no evidence was found for a change in $\psi$ at either radio or optical wavelengths. The X-ray rotation was interpreted as a shock propagating along a helical magnetic field.

In the case of PG~1553+113, we instead have observed $\Delta\psi\approx125^\circ$ at optical wavelengths, which occurred between \ixpe\ intervals a and b. This rotation of the optical polarization angle was accompanied by a modestly variable X-ray flux and roughly constant spectral shape. Moreover, we have found no evidence for a change in $\psi$ in either the radio or X-ray band. This is consistent with the interpretation of the Mrk~421 results, where the X-ray-optical--radio emission of blazars originates not from a single localized region, but rather from separate (or at most partially co-spatial) regions in the jet.
Moreover, as imaged with the MOJAVE program\footnote{https://www.cv.nrao.edu/MOJAVE/sourcepages/1553+113.shtml}, PG1553+113 is a very core-dominated source at 15 GHz, with faint extended structure at a projected position angle between 30-50$^{\circ}$. This implies that the observed X-ray polarization position angle is oblique to the parsec-scale jet direction, with a difference $\sim45^{\circ}$, similar to that observed for Mrk~421 during its first
\ixpe~observation in 2022 May \citep[][]{DiGesu2023}. However, we note that the position angle of the jet is known to vary over time \citep{Lico2020}.

The origin of the optical rotation is unclear. Figure \ref{plt:multi_obs} (bottom panel) shows the 3-day-binned $\gamma$-ray light curve of PG~1553+113 during the \ixpe~ observations. There is a clear brightening of the $\gamma$-rays near the center of the optical polarization angle rotation. A statistical association between optical rotations and $\gamma$-ray activity has already been established by the RoboPol program \citep{Blinov2015,Blinov2018}, suggesting that such rotations are often deterministic rather than a random walk of the polarization angle. However, the statistics do not exclude the possibility that some rotations are indeed random walks.
Kink-driven magnetic reconnection \citep[][]{Bodo2021} can cause variations of the polarization angle that would be different in optical and X-rays. However, in this scenario, the optically emitting particles do not travel far from the current sheet, therefore we expect a similar polarization degree in optical and X-rays \citep{Bodo2021}. The observed chromatic behavior of $\Pi$ disfavors this interpretation. Merging of reconnection plasmoids can also produce orphan optical polarization rotations accompanied by multiwavelength flares  \citep{Hosking2020}. However, the timescales of such rotations are expected to be much shorter than the $\sim7$ days observed here. In addition, we did not observe the expected optical flare, which makes this interpretation unlikely.

We conclude that the lower degree of polarization at optical wavelengths is most likely connected  with stronger turbulence in the optical emission region. In this case, the detected rotation could have been caused by turbulence, as observed in simulated polarization angle behavior in the Turbulent Extreme Multi-Zone (TEMZ) model and Particle-in-Cell simulations \citep{Marscher2017,Marscher2021,Zhang2023}.

Given the above assessment, our results favor a model where an energy stratified --- and most likely shock-accelerated --- electron population is responsible for the non-thermal X-ray--optical--radio emission of at least some, and possibly all, astrophysical jets associated with SMBHs.

%\begin{acknowledgments}
{\it \textbf{Acknowledgments:}}
The Imaging X-ray Polarimetry Explorer (\ixpe) is a joint US and Italian mission.  The US contribution is supported by the National Aeronautics and Space Administration (NASA) and led and managed by its Marshall Space Flight Center (MSFC), with industry partner Ball Aerospace (contract NNM15AA18C).  The Italian contribution is supported by the Italian Space Agency (Agenzia Spaziale Italiana, ASI) through contract ASI-OHBI-2017-12-I.0, with agreements ASI-INAF-2022-14-HH.0 and ASI-INFN 2021-43-HH.0, and its Space Science Data Center (SSDC), and by the Istituto Nazionale di Astrofisica (INAF) and the Istituto Nazionale di Fisica Nucleare (INFN) in Italy.  This research used data products provided by the IXPE Team (MSFC, SSDC, INAF, and INFN) and distributed with additional software tools by the High-Energy Astrophysics Science Archive Research Center (HEASARC), at NASA Goddard Space Flight Center (GSFC). We acknowledge financial support from ASI-INAF agreement n.\ 2022-14-HH.0. The research at Boston University was supported in part by National Science Foundation grant AST-2108622, NASA Fermi Guest Investigator grants 80NSSC21K1917 and 80NSSC22K1571, and NASA {\rm Swift} Guest Investigator grant 80NSSC22K1482. This work was supported in part by NSF grant AST-2109127.
This study was based in part on observations conducted using the Perkins Telescope Observatory (PTO) in Arizona, USA, which is owned and operated by Boston University. We acknowledge the use of public data from the Swift data archive. Based on observations obtained with XMM-Newton, an ESA science mission with instruments and contributions directly funded by ESA Member States and NASA. We acknowledge funding to support our NOT observations from the Finnish Centre for Astronomy with ESO (FINCA), University of Turku, Finland (Academy of Finland grant nr 306531). The IAA-CSIC co-authors acknowledge financial support from the Spanish "Ministerio de Ciencia e Innovaci\'{o}n" (MCIN/AEI/ 10.13039/501100011033) through the Center of Excellence Severo Ochoa award for the Instituto de Astrof\'{i}sica de Andaluc\'{i}a-CSIC (CEX2021-001131-S), and through grants PID2019-107847RB-C44 and PID2022-139117NB-C44. The POLAMI observations were carried out at the IRAM 30m Telescope. IRAM is supported by INSU/CNRS (France), MPG (Germany) and IGN (Spain). Part of the French contribution is supported by the Scientific Research National Center (CNRS) and the French Space Agency (CNES). Some of the data are based on observations collected at the Observatorio de Sierra Nevada, owned and operated by the Instituto de Astrof\'{i}sica de Andaluc\'{i}a (IAA-CSIC). Further data are based on observations collected at the Centro Astron\'{o}mico Hispano-Alem\'{a}n(CAHA), operated jointly by Junta de Andaluc\'{i}a and Consejo Superior de Investigaciones Cient\'{i}ficas (IAA-CSIC).  CC acknowledges support by the European Research Council (ERC) under the HORIZON ERC Grants 2021 programme under grant agreement No. 101040021. The Submillimetre Array is a joint project between the Smithsonian Astrophysical Observatory and the Academia Sinica Institute of Astronomy and Astrophysics and is funded by the Smithsonian Institution and the Academia Sinica. Mauna Kea, the location of the SMA, is a culturally important site for the indigenous Hawaiian people; we are privileged to study the cosmos from its summit. Some of the data reported here are based on observations made with the Nordic Optical Telescope, owned in collaboration with the University of Turku and Aarhus University, and operated jointly by Aarhus University, the University of Turku, and the University of Oslo, representing Denmark, Finland, and Norway, the University of Iceland and Stockholm University at the Observatorio del Roque de los Muchachos, La Palma, Spain, of the Instituto de Astrofisica de Canarias. E. L. was supported by Academy of Finland projects 317636 and 320045. The data presented here were obtained in part with ALFOSC, which is provided by the Instituto de Astrofisica de Andalucia (IAA) under a joint agreement with the University of Copenhagen and NOT.  S. Kang, S.-S. Lee, W. Y. Cheong, S.-H. Kim, and H.-W. Jeong  were supported by the National Research Foundation of Korea (NRF) grant funded by the Korea government (MIST) (2020R1A2C2009003). The KVN is a facility operated by the Korea Astronomy and Space Science Institute. The KVN operations are supported by KREONET (Korea Research Environment Open NETwork) which is managed and operated by KISTI (Korea Institute of Science and Technology Information). Partly based on observations with the 100-m telescope of the MPIfR (Max-Planck-Institut f\"ur Radioastronomie) at Effelsberg. Observations with the 100-m radio telescope at Effelsberg have received funding from the European Union’s Horizon 2020 research and innovation programme under grant agreement No 101004719 (ORP).

%\end{acknowledgments}

\facilities{Calar Alto, Effelsberg 100-m, {\rm Fermi}, IRAM-30m, \ixpe, KVN, NOT, Perkins, SMA, Swift, T90, XMM-Newton}

\bibliography{biblio}
\bibliographystyle{aasjournal}

\end{document}